\documentclass[aps,prl,twocolumn,superscriptaddress]{revtex4}
\pdfoutput=1

\usepackage{graphicx}
\usepackage{amsmath}
\usepackage{amssymb}
\usepackage{bm}

\DeclareMathOperator{\tr}{tr}

\begin{document}
\title{Excitable Patterns in Active Nematics}

\author{L. Giomi}
\affiliation{School of Engineering and Applied Sciences, Harvard University, Cambridge, MA 02138,  USA}
\affiliation{Martin A. Fisher School of Physics, Brandeis University, Waltham, MA 02454, USA}

\author{L. Mahadevan}
\affiliation{School of Engineering and Applied Sciences, Harvard University, Cambridge, MA 02138,  USA}

\author{B. Chakraborty}
\affiliation{Martin A. Fisher School of Physics, Brandeis University, Waltham, MA 02454, USA}

\author{M. F. Hagan}
\affiliation{Martin A. Fisher School of Physics, Brandeis University, Waltham, MA 02454, USA}

\date{\today}

\begin{abstract}
\noindent We analyze a model of mutually-propelled filaments suspended in a two-dimensional solvent. The system
undergoes a mean-field isotropic-nematic transition for large enough filament concentrations and the nematic order
parameter is allowed to vary in space and time. We show that the interplay between non-uniform nematic order,
activity and flow results in spatially modulated relaxation oscillations, similar to those seen in excitable media.
In this regime the dynamics consists of nearly stationary periods separated by ``bursts'' of activity in which the
system is elastically distorted and solvent is pumped throughout. At even higher activity the dynamics becomes
chaotic.

\end{abstract}

\maketitle
\noindent{Colonies of motile microorganisms, the cytoskeleton and its components, cells and tissues have much in
common with soft condensed matter systems (i.e. liquid crystals, amphiphiles, colloids etc.), but also exhibit
phenomena that do not appear in inanimate matter. These unique properties arise when the constituent particles are
active: they consume and dissipate energy to fuel internal changes that generally lead to motion. When active
particles have elongated shapes, as seen in cytoskeletal filaments and some cells, they undergo orientational
ordering at high concentration to form liquid crystalline phases. The theoretical and experimental study of active
materials has disclosed a wealth of emergent behaviors, such as the occurrence of \emph{giant} density fluctuations
\cite{GiantFluctuations}, the emergence of spontaneously flowing states (sometimes referred to as flocking)
\cite{SpontaneousFlow}, unconventional rheological properties \cite{Rheology} and new spatiotemporal patterns not
seen in passive complex fluids \cite{PatternFormation}. Recently Schaller {\em et al}. \cite{Schaller:2010} observed
collective motion, large-scale fluctuations in density and the degree of polar order, and swirling motions in a
motility assay consisting of highly concentrated actin filaments propelled by immobilized molecular motors in a
planar geometry. However, the range of behaviors realizable in active materials and the connection between material
characteristics and system dynamics remain incompletely understood. In this letter, we show that active systems
exhibit behaviors similar to those of excitable systems, showing relaxation oscillations that couple activity to
spontaneous pulsatile flow with quiescent periods in between, similar to biological pumps. We furthermore show that
systems which undergo nematic ordering can give rise to large-scale swirling motions resembling those observed in
Ref.~\cite{Schaller:2010} even in the absence of polar order. These behaviors were not seen in previous
investigations because they arise only when the degree of nematic order is allowed to fluctuate in space and time.

Our model consists of mutually-propelled elongated particles suspended in a solvent confined to two dimensions, as
seen for example in recent motor-filament assays \cite{Schaller:2010}. The dynamical variables in such a system are
particle concentrations $c$, the solvent flow field {\bf v} and the nematic tensor
$Q_{ij}=S(n_{i}n_{j}-\frac{1}{2}\delta_{ij})$, with $S$ the nematic order parameter and {\bf n} the director field,
all of which are allowed to vary in space and time. We assume that the suspension of rod-like active particles have
length $\ell$ and mass $M$ in a solvent of density $\rho_{\rm solvent}$. The total density of the system
$\rho=Mc+\rho_{\rm solvent}$ is conserved, so the fluid is assumed to be incompressible. The total number of active
particles is also constant, thus the concentration $c$ obeys the continuity equation:
\begin{equation}\label{eq:continuity-equation}
\partial_{t}c = -\nabla\cdot({\bf j}^{p}+{\bf j}^{a})\,,
\end{equation}
where ${\bf j}^{p}$ and ${\bf j}^{a}$ are respectively the passive and active contributions to the
current density. The passive current density has the standard form $j_{i}^{p}=cv_{i}-D_{ij}\partial_{j}c$ where
$D_{ij} = D_{0}\delta_{ij}+D_{1}Q_{ij}$ is the anisotropic diffusion tensor, while the active current can be
constructed phenomenologically to be of the form $j_{i}^{a}=-\alpha_{1}c^{2}\partial_{j}Q_{ij}$ or derived from
microscopic models \cite{Ahmadi:2006}. Here the factor $c^2$ reflects the fact that activity arises from
interactions between pairs of rods while the constant $\alpha_{1}$ describes the level of activity and is
proportional to the concentration of motors and the rate of adenosine-triphosphate (ATP) consumption.

The flow velocity obeys an active form of the Navier-Stokes equation
\begin{equation}\label{eq:navier-stokes}
\rho\partial_{t}v_{i} = \eta\Delta v_{i} -\partial_{i}p+\partial_{j}\tau_{ij}\,,
\end{equation}
with $\eta$ the viscosity, $p$ the pressure, and the active stress tensor $\tau_{ij}$  given by:
\begin{equation}\label{eq:stress}
\tau_{ij}=-\lambda S H_{ij}+Q_{ik}H_{kj}-H_{ik}Q_{kj}+\alpha_{2}c^{2}Q_{ij}
\end{equation}
Here $\nabla \cdot v = 0$, and the first three terms in Eq. \eqref{eq:stress} represent the elastic stress due to
the liquid crystalline nature of the system, with $H_{ij}=-\delta F/\delta Q_{ij}$ the molecular tensor defined from
the two-dimensional Landau-De Gennes free energy:
\[
F/K = \int
dA\,[\tfrac{1}{2}(\nabla\cdot\bm{Q})^{2}+\tfrac{1}{4}(c^{*}-c)\,\tr\bm{Q}^{2}+\tfrac{1}{4}c\,(\tr\bm{Q}^{2})^{2}]
\]
where $K$ is the splay and bending stiffness in the one-constant approximation. At equilibrium, above the critical
concentration $c^{*}$,  $S=\sqrt{2\tr\bm{Q}^{2}}=\sqrt{1-c^{*}/c}$  consistent with hard-rod fluid models where the
isotropic-nematic (IN) transition is driven only by the concentration of the nematogens. The last term was first
introduced in Ref. \cite{Pedley:1992} and represents the tensile/contractile stress exerted by the active particles
in the direction of the director field ${\bf n}$ with $\alpha_{2}$ a second activity constant.

Finally, the nematic order parameter $Q_{ij}$ satisfies a hydrodynamic equation that can be obtained by constructing
all possible traceless-symmetric combinations of the relevant fields, namely the strain-rate tensor
$u_{ij}=\frac{1}{2}(\partial_{i}v_{j}+\partial_{j}v_{i})$, the vorticity tensor
$\omega_{ij}=\frac{1}{2}(\partial_{i}v_{j}-\partial_{j}v_{i})$ and the molecular tensor $H_{ij}$
\cite{Olmsted:1992}, so that
\begin{equation}\label{eq:oseen}
[\partial_{t}+{\bf v}\cdot\nabla]Q_{ij} = \gamma^{-1}H_{ij} + \lambda S u_{ij}+Q_{ik}\omega_{kj}-\omega_{ik}Q_{kj}
\end{equation}
where $\gamma$ is an orientational viscosity, and the additional terms on the right-hand side describe the coupling
between nematic order and flow in two dimensions, with $\lambda$ the flow-alignment parameter which dictates how the
director field rotates in a shear flow and affects the flow and rheology of active systems
\cite{SpontaneousFlow,Rheology}.

\begin{figure}[t]
\includegraphics[width=.9\columnwidth]{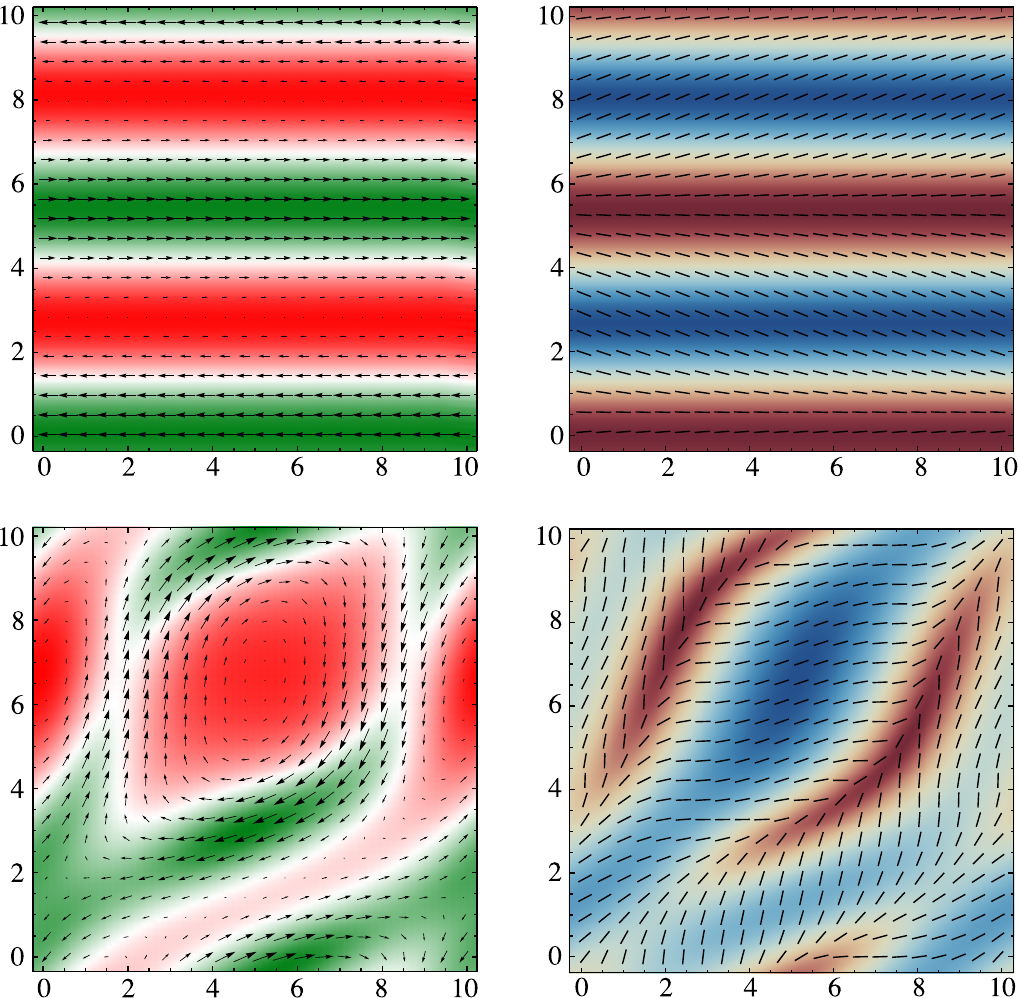}
\caption{\label{fig:flows} (Color online) The velocity filed (left) and the director field (right) superimposed to a
density plot of the concentration and the nematic order parameter for $\alpha_{2}=0.4$ (top) and $\alpha_{2}=3$
(bottom). The colors indicate regions of large (green) and small (red) density and large (blue) and small (brown)
nematic order parameter. For moderate values of $\alpha_{2}$ the flow consists of two bands traveling in opposite
directions with the director field is nearly uniform inside each band. For large $\alpha_{2}$ the flow is
characterized by large vortices that span lengths of the order of the system size and the director field is
organized in grains.}
\end{figure}

The dynamics of such an active nematic suspension is governed by the interplay between the active forcing, whose
rate $\tau_{a}^{-1}$ is proportional to the activity parameters $\alpha_{1}$ and $\alpha_{2}$, and the relaxation of
the passive structures, the solvent and the nematic phase, in which energy is dissipated or stored. The response of
the passive structures, as described here, occurs at three different time scales: the relaxational time scale of the
nematic degrees of freedom $\ell^{2}/(\gamma^{-1}K)$, the diffusive time scale $\ell^{2}/D_{0}$, the dissipation
time scale of the solvent $\rho L^{2}/\eta$, where $L$ is the system size. While the presence of three dimensionless
parameters makes for a very rich phenomenology, we temporarily assume that the three passive time scales are of the
same magnitude $\tau_{p}$. When $\tau_{a}\gg\tau_{p}$, the active forcing is irrelevant and the system is just a
traditional passive suspension. On the other hand, when $\tau_{a}\sim\tau_{p}$, the passive structures can balance
the active forcing leading to a stationary regime in which active stresses are accommodated via both elastic
distortion and flow. Finally, when $\tau_{a}\ll\tau_{p}$ the passive structures will fail to keep up, leading to a
dynamical and possibly chaotic interplay between activity, nematic order and flow. In the rest of the paper, we
quantify these different regimes.  We first make the system dimensionless by scaling all lengths using the rod
length $\ell$, scaling time with the relaxation time of the director field $\tau_{p}=\ell^{2}/(\gamma^{-1}K)$ and
scaling stress  using the elastic stress $\sigma=K\ell^{-2}$.

We start with a linear stability analysis of the hydrodynamic equations about the homogeneous solution letting
$\bm{\varphi}=\{c,\,Q_{xx},\,Q_{xy},\,2\omega_{xy}\}=\bm{\varphi}_0+\delta \bm{\varphi}(t)$ with
$\bm{\varphi}_{0}=\{c_{0},\,S_{0}/2,\,0,\,0\}$, and consider solutions with periodic boundary conditions on a square
domain of the form $\delta \bm{\varphi}({\bf x},t) =
\sum_{n=-\infty}^{\infty}\sum_{m=-\infty}^{\infty}\bm{\varphi}_{nm}(t)\exp[\frac{2\pi i}{L}(nx+my)]$. With this
choice the linearized hydrodynamic equations reduce to a set of coupled of linear ordinary differential equations
for the Fourier modes $\bm{\varphi}_{nm}$: $\partial_{t}\bm{\varphi}_{nm} = {\bm A}_{nm}\bm{\varphi}_{nm}$. The
first mode to become unstable is the transverse excitation $(n,\,m)=(0,\,1)$ associated with the block-diagonal
matrix:
\[
{\bm A}_{01} =
\left(
\begin{array}{cc}
{\bf a}_{01} & 0 \\ 0 & {\bf b}_{01}
\end{array}
\right)\,,
\]
where:
\[
{\bm a}_{01} =
\left(
\begin{array}{cc}
-\frac{2\pi^{2}}{L^{2}}\,(2D_{0}-D_{1}S_{0}) &
 \frac{4\pi^{2}}{L^{2}}\,\alpha_{1}c_{0}^{2} \\
 \frac{c^{*}}{4c_{0}}\,S_{0} &
-c_{0}S_{0}^{2}-\frac{4\pi^{2}}{L^{2}}
\end{array}
\right)
\]
and:
\[
{\bm b}_{01} =
\left(
\begin{array}{cc}
-\frac{4\pi^{2}}{L^{2}} &
 \frac{1}{2}(1-\lambda)S_{0} \\
 \frac{4\pi^{2}}{L^{2}}\,\alpha_{2}c_{0}^{2}-\frac{16\pi^{4}}{L^{4}}\,S_{0}(1-\lambda) &
-\frac{4\pi^{2}\eta}{L^{2}}
\end{array}
\right)\,.
\]
The mechanism that triggers the instability of the homogeneous state relies on the coupling between local
orientations and flow embodied in the matrix ${\bm b}_{01}$, whose diagonalization yields a lower critical value of
$\alpha_{2}$:
\begin{equation}\label{eq:alpha_c2}
\alpha_{2}^{*} = \frac{4\pi^{2}[2\eta+S_{0}^{2}(1-\lambda)^{2}]}{c_{0}^{2}L^{2}S_{0}(1-\lambda)}\,.	
\end{equation}
We see that the critical value of the activity required for instability falls with increasing system size and
increases with the viscosity of the solvent. In general, shear flow causes the director field to rotate for
$\lambda\ne 1$, which generates elastic stress. For small activity, the elastic stiffness dominates and suppresses
flow, while above $\alpha_2^{*}$ we observe collective motion.

To go beyond this simple analysis, we numerically integrated the hydrodynamic equations on a two-dimensional
periodic domain with initial configurations of a homogeneous system whose director field was aligned along the $x$
axis subject to a small random perturbation in density and orientation, with $\alpha_{1}=\alpha_{2}/2$,
$\eta=c^{*}=D_{0}=D_{1}=1$, $\lambda=0.1$ and $L=10$. As predicted by the linear stability analysis, at low activity the
system relaxes to a stationary homogeneous state with $v_{x}=v_{y}=0$ and $S = \sqrt{1-c^{*}/c}$. Above  the
critical value $\alpha_{2}^{*}$, the system forms two bands flowing in opposite directions. The solution is constant
along the flow direction (see Fig. \ref{fig:flows} top) while the direction of the streamlines (in this case along
the $x$ direction) is dictated by the initial conditions. As shown in Fig. \ref{fig:laminar}, the optima in the flow
velocity correspond to the maximal distortion of the director field ${\bf n}$. Variations in concentration $c$ and
the nematic order parameter $S$ are of order $2\%$ with a minimum in $S$ at the center of a flowing band due to the
balance between diffusive and active currents. We note that these density modulations parallel to the nematic
director are characteristic of systems with nematic order, while the density bands orthogonal to the direction of
alignment observed in Ref.~\cite{Schaller:2010} are consistent with polar order \cite{SpontaneousFlow}.

\begin{figure}[t]
\includegraphics[width=.9\columnwidth]{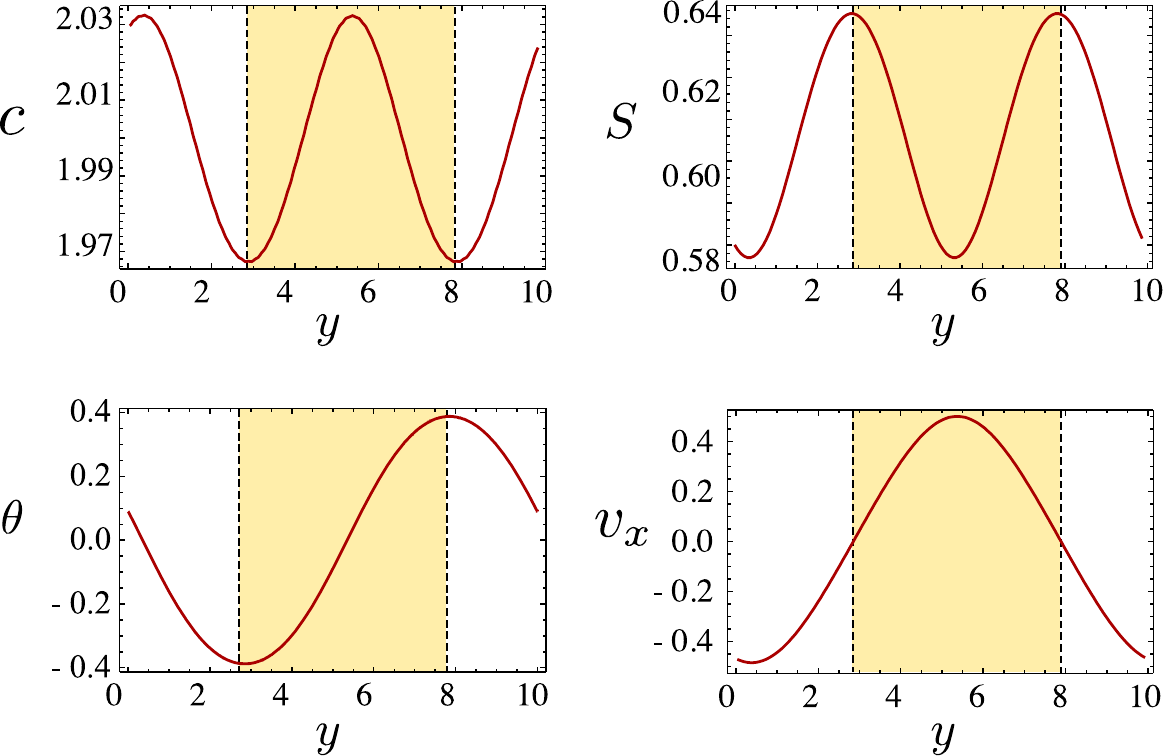}
\caption{\label{fig:laminar}(Color online) The hydrodynamic fields $c$, $S$, $\theta$ and $v_{x}$ along the $y$ axis
for $\alpha_{2}=0.4$ The yellow region indicates the band visible in the top panels of Fig. \ref{fig:flows}}
\end{figure}

\begin{figure}
\centering
\includegraphics[width=.9\columnwidth]{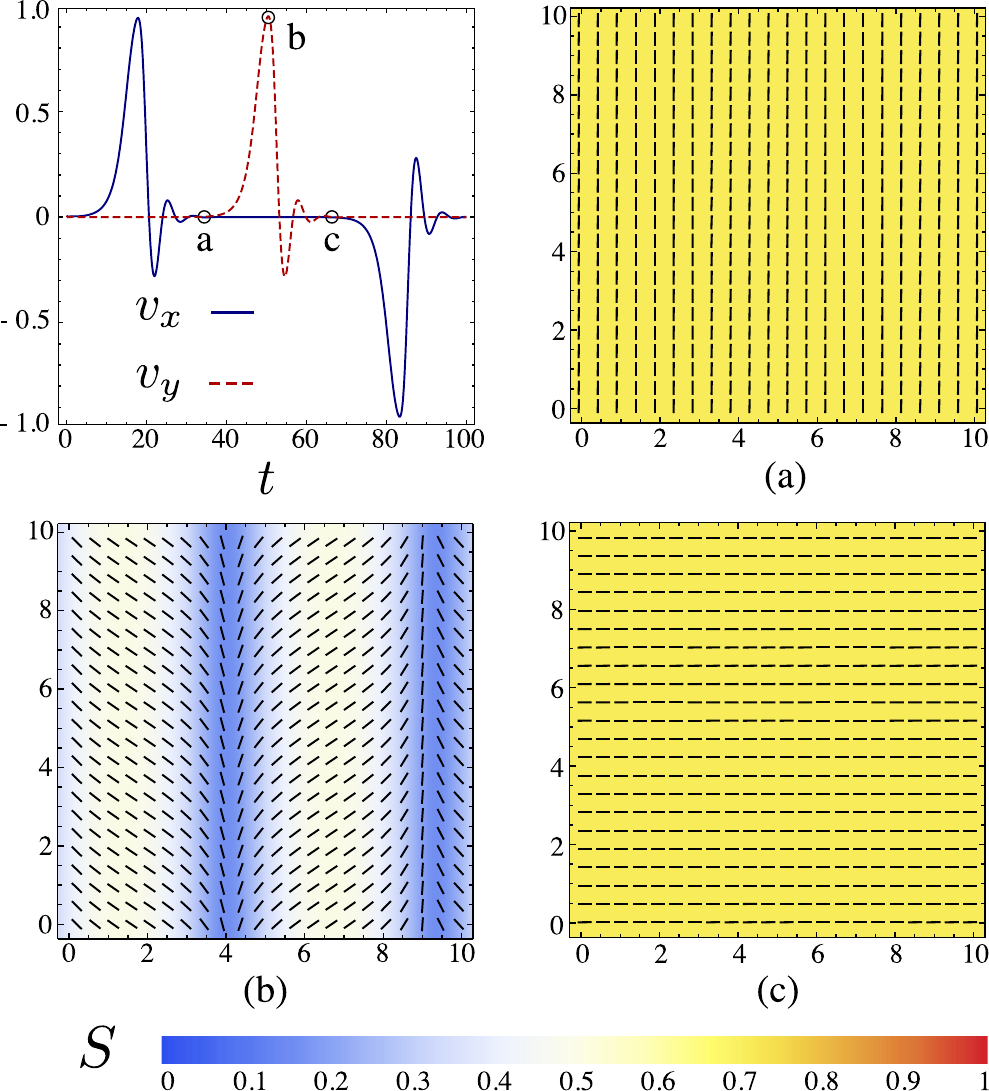}
\caption{\label{fig:burst}(Color online) Dynamics of active ``burst'' for $\alpha_{2}=1.5$. The flow velocity at the
point $x=y=L/3$ is shown as a function of time over the course of a director field rotation (top left) and the
director field is shown for the three labeled time points. Between two consecutive bursts the system is homogeneous
and uniformly aligned. During a burst, nematic order is drastically reduced in the whole system and the director
undergoes a distortion with a consequent formation of two bands flowing in opposite directions. After a burst, a
stationary state is restored with the director field rotated of  $90^{\circ}$ with respect to its previous
orientation.}		
\end{figure}	

Upon increasing the value of the activity parameter $\alpha_{2}$, the spontaneously flowing state evolves into a
pulsatile spatial relaxation oscillator. Fig. \ref{fig:burst} (top left panel) shows a plot of the $x$ and $y$
component of the flow velocity in the center of the box for $\alpha_{2}=1.5$. In this regime the dynamics consists
of a sequence of almost stationary passive periods separated by active ``bursts'' in which the director switches
abruptly between two orientations. During passive periods, $c$ and $S$ are nearly uniform, there is virtually no
flow and the director field is either parallel or perpendicular to the $x$ direction. Eventually this configuration
breaks down and the director field rotates by $90^{\circ}$ (see Fig. \ref{fig:burst}). The rotation of the director
field is initially localized along lines, generating a band of flow  similar to those in Fig. \ref{fig:flows} (top).
The flow terminates after the director field rotates and a uniform orientation is restored. The process then
repeats.

This rotation of the director field occurs through a temporary ``melting'' of the nematic phase. As shown in Fig.
\ref{fig:burst}, during each passive period the nematic order parameter is equal to its equilibrium value
$S_{0}=\sqrt{1-c^{*}/c}$, but drops to $\sim \frac{2}{5}S_{0}$ during rotation.  The reduction of order is system-wide,
but, as shown in the middle in the bottom-left panel of Fig. \ref{fig:burst}, is most pronounced along the
boundaries between bands. Without transient melting, the distortions of the director field required for a burst are
unfavorable for any level of activity.

\begin{figure}[t]
\centering
\includegraphics[width=1.\columnwidth]{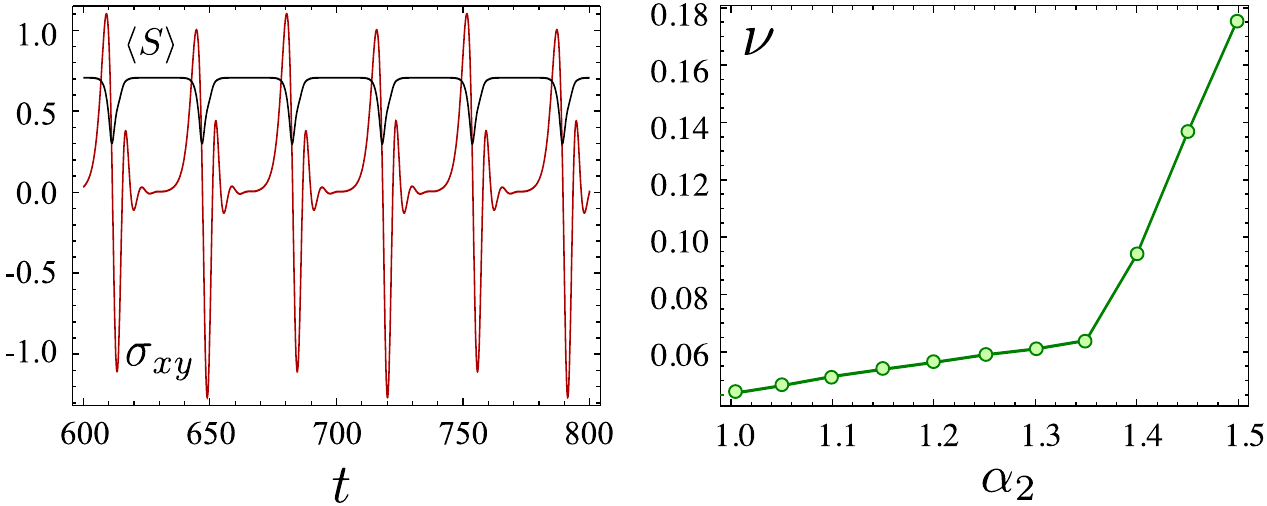}
\caption{\label{fig:stress}(Color online)  (Left) The average nematic order parameter $\langle S\rangle$ and the
total shear stress $\sigma_{xy}$ are shown over several bursts for $\alpha_{2}=1.5$. (right) The frequency of bursts
is shown as a function of $\alpha_2$.}
\end{figure}

To illustrate the origin of the relaxation oscillations it is useful to construct a simplified version of the
hydrodynamic equations by retaining the minimal features responsible
for the oscillatory phenomenon: the coupling between active forcing and the fluid microstructure and the variable
nematic order embedded in the Landau-De Gennes free energy. This can be achieved by approximating $Q_{xx}^{2}$ as
a constant and $u_{xx}\approx 0$. Then, letting $u=-u_{xy}$ and $Q=Q_{xy}$, and dropping the coupling between $Q_{ij}$
and $\omega_{ij}$, Eqs. \eqref{eq:navier-stokes} and \eqref{eq:oseen} can be expressed in Fourier space as:
\begin{subequations}\label{eq:van-der-pol}
\begin{gather}
\dot{Q} = aQ-bQ^{3}-u \\[5pt]
\dot{u} = k^{2}(\alpha Q-u)
\end{gather}
\end{subequations}
where $k$ is a wave number of an arbitrary spatial mode and $a$ and $b$ are constants. Eqs. \eqref{eq:van-der-pol}
has the form of the FitzHugh-Nagumo model for excitable dynamical systems. For $\alpha<\frac{1}{3}(2a+k^{2})$ the
system rapidly relaxes to a state characterized by a finite strain-rate that balances the active stress: $u=\alpha Q
= \alpha\sqrt{(a-\alpha)/b}$. For $\alpha>\frac{1}{3}(2a+k^{2})$, this state becomes unstable and the trajectory
converges to a limit cycle with a frequency $\nu \sim k^{2}\alpha$. As anticipated, when the active and passive
time-scales are comparable the active forcing is accommodated by the microstructure leading to a distortion of the
director field and a steady flow. However, when the active forcing rate is increased, the microstructure dynamics
lag, resulting in oscillatory dynamics. Clearly, the full system exhibits a much richer behavior than that captured
by Eqs. \eqref{eq:van-der-pol}, but the qualitative trends persist. For instance, the increase in the slope of $\nu$
shown in Fig. \ref{fig:stress} is not simply associated with the excitation of a second spatial mode of larger
wave-number and is more likely due to the dynamics of the concentration field, which has been ignored in Eqs.
\eqref{eq:van-der-pol}.

When the activity $\alpha_{2}$ is further increased, the sequence of low activity periods and bursts becomes more
complex and eventually chaotic. We emphasize that the oscillatory dynamics and the chaotic flows discussed below do
not occur in our equations unless the magnitude of the nematic order parameter is allowed to fluctuate. Fig.
\ref{fig:flows} (bottom) shows a typical snapshot of the flow velocity and the director field superimposed to a
density plot of the concentration and the nematic order parameter respectively. The flow is characterized by large
vortices with positions related to ``grains'' in which the director field is uniformly oriented. The grain
boundaries are of the order of the system size and are the fastest flowing regions in the system. Thus the dynamics
in this regime is characterized by sets of grains of approximatively uniform orientation that swirl around each
other and continuously merge and reform.

In summary, we analyzed the hydrodynamics of active nematic suspensions in two dimensions. By allowing spatial and
temporal fluctuations in the nematic order parameter we have observed a rich interplay between order, activity and
flow. Significantly, we find that allowing fluctuations in the magnitude of the order parameter $S$ qualitatively
changes the flow behavior as compared to systems in which $S$ is constrained to be uniform. Finally, we note
that both the flip-flop dynamics (Fig.~\ref{fig:burst})  and the swirling motion (Fig.~\ref{fig:flows} bottom)
resemble behavior observed in the motility assay experiments of Schaller {\em et al}. \cite{Schaller:2010}. Our
analysis suggests that these classes of patterns can emerge even in the absence of polar order.

\acknowledgments
We gratefully acknowledge support from the NSF Brandeis MRSEC (LG, BC, and MFH), the NSF Harvard MRSEC (LG,LM),
NIAID R01AI080791 (MFH), the Harvard Kavli Institute for Nanobio Science \& Technology (LG, LM), and the MacArthur
Foundation (LM).


\begin{thebibliography}{1}

\bibitem{GiantFluctuations}
V. Narayan, S. Ramaswamy and N. Menon, Science {\bf 317}, 105 (2007). J. Deseigne, O. Dauchot and H. Chat\'e, Phys.
Rev. Lett. {\bf 105}, 098001 (2010).

\bibitem{SpontaneousFlow}
R. Voituriez, J. F. Joanny and J. Prost, Europhys. Lett. \textbf{70}, 118102 (2005). D. Marenduzzo, E. Orlandini, M.
E. Cates, and J. M. Yeomans, Phys. Rev. E {\bf 76}, 031921 (2007). L. Giomi, M. C. Marchetti and T. B. Liverpool,
Phys. Rev. Lett. {\bf 101}, 198101 (2008). S. Mishra, A. Baskaran and M. C. Marchetti, Phys. Rev. E {\bf 81}, 061916
(2010).

\bibitem{Rheology}
M. E. Cates, S. M. Fielding, D. Marenduzzo, E. Orlandini and J. M. Yeomans, Phys. Rev. Lett. {\bf 101}, 068102
(2008). A. Sokolov and I. S. Aranson, Phys. Rev. Lett. {\bf 103}, 148101 (2009). L. Giomi, T. B. Liverpool and M. C.
Marchetti, Phys. Rev. E {\bf 81}, 051908 (2010). D. Saintillan, Phys. Rev. E {\bf 81}, 056307 (2010).

\bibitem{PatternFormation}
D. Saintillan and M. J. Shelley, Phys. Rev. Lett. {\bf 100}, 178103 (2008). H. Chat\'e, F. Ginelli and R. Montagne,
Phys. Rev. Lett. {\bf 96}, 180602 (2006). F. Ginelli \emph{et al.} Phys. Rev. Lett. {\bf 104}, 184502 (2010)

\bibitem{Schaller:2010}
V. Schaller, C. Weber, C. Semmrich, E. Frey and A. R. Bausch, Nature {\bf 467}, 73 (2010).

\bibitem{Ahmadi:2006}
A. Ahmadi, M. C. Marchetti and T. B. Liverpool, Phys. Rev. E {\bf 74}, 061913 (2006). T. B. Liverpool and M. C.
Marchetti, {\em Hydrodynamics and rheology of active polar filaments}, in Cell Motility, P. Lenz ed. (Springer, New
York, 2007).

\bibitem{Olmsted:1992}
P. D. Olmsted and P. M. Goldbart, Phys. Rev. A {\bf 46}, 4966 (1992).

\bibitem{Pedley:1992}
T. J. Pedley and J. O. Kessler, Annu. Rev. Fluid Mech. {\bf 24}, 313 (1992).



\end{thebibliography}
\end{document}